\documentclass[11pt,twoside]{article}

\usepackage{asp2006}
\usepackage{epsf}
\usepackage{psfig}
\usepackage{lscape}
\usepackage{natbib}

\begin{document}
\title{How ubiquitous are massive starbursts in interacting galaxies?}   %%% Fill in title
\author{P. Di Matteo}   %%% Fill in author names
\affil{Observatoire de Paris, section de Meudon, GEPI, 5 Place Jules Jannsen, 92195 Meudon, France}    %%% Fill in author affiliations
\author{F. Bournaud, M. Martig}   %%% Fill in author names
\affil{CEA Saclay, IRFU, SAp, 91190 Gif-sur-Yvette, France}
\author{F. Combes, A.-L. Melchior, B. Semelin}   %%% Fill in author names
\affil{Observatoire de Paris, LERMA, 61 Avenue de l'Observatoire, 75014 Paris, France}

\begin{abstract} 
Many evidences exist for a connection between galaxy interactions and induced star formation. However, a large range of responses of galaxies to tidal interactions is found, both in observations and in numerical simulations. We will discuss some recent results obtained analysing a large sample ($\sim 1000$) of simulations of interacting pairs and their agreement with the most recent observational works.
\end{abstract}

%%% MAIN BODY OF TEXT GOES HERE. CONSULT "INSTRUCTIONS FOR AUTHORS USING
%%% LATEX2E MARKUP", SECTIONS 2.3-2.6 FOR HELP WITH EQUATIONS, FIGURES,
%%% AND TABLES.

\section{Introduction}   %%% Top level section head (remove "%" symbol)
\subsection{Interactions and star formation: observations} 
In the local Universe, the response of galaxies to mutual interactions is quite varied.
The strongest starbursts are found in interacting and merging systems \citep{jowr85, sandetal88, clemetal96, sandmir96, scovetal00, arribetal04, alonsoetal06} and many studies have shown that interacting pairs show an increased star formation, lasting few $10^7-10^8$ years (see Kennicutt et al. 1996 for a review), which can take place in the galaxy centers, as it is the case of M˜82 \citep{degr01}, or in the overlapping regions between galaxies, as for the Antennae \citep{wangetal04}. Nevertheless, interactions and mergers do not seem a sufficient condition to trigger high star formation episodes.  \citet{bergetal04} have shown only a weak enhancement in the star formation rate (hereafter SFR) of a sample of interacting pairs (by a factor of 2-3 in their centers), when comparing their optical colors to those of a sample of non-interacting systems. More recently, \citet{knapjam09}  have shown that the SFR of galaxies with close companions is enhanced by a factor of around two when compared to galaxies without companions and that the increase in the SFR does not occur in a burst mode of massive star formation, but it is rather continuous, with a duration of the order $10^8-10^9$ years.\\
At higher redshifts, merger-induced star formation is debated too.  \citet{consetal03} suggested that about two thirds of submillimiter galaxies at $z>1$ are undergoing a major merger; also \citet{bridetal07} argued for a significant role of mergers in the SFR density at $z\sim 1$. However, \citet{belletal05} found that less than one third of actively star forming galaxies at  $z\sim 0.8$ are interacting systems. Similar results have been found by \citet{jogetal09}, who showed that merging systems only account for a small fraction ($<30\%$) of the cosmic SFR between $0.24\le z \le 0.8$, and that at these redshifts the mean SFR of merging systems is only modestly enhanced compared to non-interacting galaxies. These results have been furtherly confirmed by   \citet{robetal09}, who argue that less than $10\%$ of star formation in massive ($M_* > 10^{10}M_{\odot}$) galaxies is triggered by mergers, thus deducing that these events do not strongly contribute to the build-up of the stellar mass since z=1.

\subsection{Interactions and star formation: numerical simulations} 
Since the work by   \citet{barnhern91}, it has been shown that tidal interactions can drive large quantities of gas into the galaxy central regions, with a subsequent increase in the SFR of the system. Negative gravity torques acting on the gas component inside the disk corotation radius  reduce its angular momentum and can cause an important inflow in the nuclear regions, where a starburst can occur  \citep{mihern94, mihern96, spri00, coxetal06, coxetal08}. The timing and the strength of the induced starburst depend on a number of parameters, in particular on the morphology of the interacting galaxies, bulgeless disk galaxies being more prone to bar instability in the first phases of the interaction, when a strong burst of star formation can occur, while bulge-disk galaxies usually show the peak of their star formation activity in the final merging phases  \citep{mihern94}. Equal-mass mergers are the most efficient mechanisms to trigger strong starbursts, and the star formation activity declines rapidly with increasing mass ratios  \citep{coxetal08}. In particular, for unequal mass mergers, while the most massive galaxy contributes to the majority of the star formation of the system, the less massive satellite usually experiences the largest enhancement in its star formation, which is hidden in the total rate of the pair (see Fig.˜8, Cox et al. 2008). \\
That galaxy interactions can lead to strong bursts of star formation, as shown in  \citet{mihern94}, does not necessarily implies that they are a sufficient condition to systematically produce high star formation enhancements. For example,  \citet{kapetal05} showed, by analysing a sampe of $\sim 50$ simulations of galaxy interactions, that the integrated SFR during an interaction is
moderately increased, up to a factor of 5 but on average a factor
of 2 with respect to that of isolated galaxies. Modelling the interacting pair NGC˜4676 (the Mice),  \citet{barnes04} showed that, independently on the numerical recipe adopted for star formation (density or shock dependent), the maximum star formation enhancement found during the interaction is a factor of ten at most with respect to pre-interaction levels. More recently,   \citet{paola07} analysed a sample of more than two hundred realisations of galaxy collisions, pointing out the difficulty to obtain strong merger-driven starbursts, and discussing the dependence of the final SFR on a number of encounter parameters (relative distance at first pericenter passage, relative velocity, strength of tidal effects, etc...).\\
In what follows, we will discuss the main results recently obtained  \citep{paola08}  by analysing two independent numerical samples of major galaxy mergers, realised employing  different numerical techniques to model baryonic and dark matter evolution, and to implement star formation. 

\section{Results from  two independent large numerical samples of major galaxy mergers} 
In  \citet{paola08} we have extended the work presented in  \citet{paola07}, by realising and analysing a sample of $\sim 890$ simulations of interacting and merging galaxies (864 for local interactions and 24 for high-redshift systems). These simulations cover a large range in morphologies (from bulgeless to early-type galaxies), orbital parameters and gas fraction (from $10\%$ to $50\%$). They have been run adopting a Tree-SPH code  \citep{ben02}, with a density-dependent star formation prescription, feedback from SNe explosions and metal enrichment. These simulations have been realised in the framework of the GalMer project and are available at  \emph{http://galmer.obspm.fr} (see also Chilingarian et al. 2009).\\
To study whether our conclusions depend on the numerical techniques and star formation schemes adopted, we have run a second set of 96 simulations with a different numerical code, a
particle-mesh code with a sticky-particle (PM-SP) modeling of the ISM (see Bournaud \& Combes 2003),
 employing also star formation models that differ from the density-dependent Schmidt law.  
\subsection{Intensity of starbursts episodes}   %%% Second level section head (remove "%" symbol)
These simulations confirm our previous results  \citep{paola07} about the variety of star formation evolutions and enhancements that can be found in interacting pairs.\\
In agreement with observations, it results that  \emph{mergers do not always trigger starbursts}, and this result appears to be robust because it is found in both numerical codes used (Tree-SPH and PM-SP) and does not change significantly when using star formation prescriptions different from a pure Schmidt law. For galaxies with gas content typical of local or
low-redshift galaxies, the starbursts induced by galaxy major mergers have a
moderate intensity, SFRs being rarely enhanced by
factors larger than 5 compared to isolated galaxies, even at the
peak of the starbursts. About $10\%$ of pairs in our sample shows a star formation enhancement at least ten times higher than the SFR of an isolated reference sample. Independently of
the code used and star formation recipe adopted, we find a median value for the maximal relative SFR of about 3.
\subsection{Duration of starbursts episodes} 
In our models, the duration of the moderate starbursts is generally smaller than 500 Myr, and no more than $15\%$ of interaction-induced
starbursts have a duration greater than 500 Myr.
\subsection{From the local Universe to higher redshifts} 
When disks have higher gas fractions (about $50\%$), interacting pairs show higher star formation rates than low-redshift systems. When normalized to the SFR of the corresponding galaxies evolved isolated, inflow-induced starbursts do not result to be  stronger than their local counterparts. Also the duration of the inflow-driven star formation enhancement is not longer than that of local systems. We note, however, that the formation of massive clumps in our gas-rich disks and their induced star formation can produce longer star formation enhancements than those found for local galaxies.\\

Strong starbursts can occur during some major mergers, as observed, but our work suggests that merger-driven gas inflow is not always an efficient process to produce strong star formation enhancements. This result seems to be consistent with a number of recent observational works  \citep{belletal05, jogetal09, robetal09} which suggest the limited role of mergers in building up the stellar mass since z=1.

%\acknowledgements %%% Text of acknowledgements runs on after this command.

%%% THE BIBLIOGRAPHY
%%%
%%% CONSULT SECTION 3 OF "INSTRUCTIONS FOR AUTHORS" FOR HOW TO USE NATBIB.
%%% AUTHORS ARE ENCOURAGED TO USE EITHER THE "THEBIBLIOGRAPY" ENVIRONMENT
%%% BY UNCOMMENTING (DELETING THE "%" SYMBOL) THE COMMANDS BELOW, OR BY
%%% USING THE BIBTEX ENVIRONMENT. TO FIND OUT WHICH IS APPLICABLE TO YOUR
%%% CONTRIBUTION, CONSULT THE VOLUME EDITORS FOR YOUR PROCEEDINGS.
%%%

\end{document}